\documentclass[twocolumn,showpacs,preprintnumbers,amsmath,amssymb]{revtex4}
\usepackage{graphicx}
\usepackage{dcolumn}
\usepackage{bm}
\usepackage[dvips]{epsfig}

\begin{document}
\title{Washboard Road:\\The dynamics of granular ripples formed by rolling wheels}

\author{Nicolas Taberlet$^{1}$}
\author{Stephen W. Morris$^{1,2}$}
\author{Jim N. McElwaine$^{1}$}

\affiliation{$^{1}$DAMTP, University of Cambridge, Wilberforce Rd, CB3 0WA Cambridge, U.-K.\\
  $^{2}$Department of Physics, University of Toronto, 60 St. George St., Toronto, Canada, M5S 1A7}

\date{\today}

\begin{abstract}   
  
  Granular surfaces tend to develop lateral ripples under the action
  of surface forces   exerted by rolling wheels, an effect known as
  \emph{washboard} or \emph{corrugated} road.
   We report the results of both laboratory experiments and soft-particle direct numerical simulations.
  Above a critical speed, the ripple pattern appears as small patches of
  traveling waves which eventually spread to the entire circumference.
  The ripples drift slowly in the driving direction.  Interesting
  secondary dynamics of the saturated ripples were observed, as well
  as various ripple creation and destruction events.  All of these
  effects are captured qualitatively by 2D soft particle simulations
  in which a disk rolls over a bed of poly-disperse particles in a
  periodic box. These simulations show that compaction and segregation are inessential to the ripple phenomenon. We also discuss a simplified scaling model which gives some insight into the mechanism of the instability.

\end{abstract}

\pacs{45.70.Qj,45.70.-n }
\maketitle

Ripples which spontaneously appear due to the action of rolling wheels
on unpaved roads bedevil transportation worldwide, especially in
developing countries~\cite{heath}.  This effect, known as
\emph{corrugated} or \emph{washboard} road, can severely limit the
usefulness of unsurfaced roads. More generally, the appearance of
ripples on a granular surface under tangential stress is reminiscent
of other sorts of wind- and water-driven ripples~\cite{langlois}, and of dune
formation~\cite{herrmann}.  This resemblance suggests that this problem, which is
well-discussed in the engineering
literature~\cite{heath,mather,riley,riley_thesis,grau,shoop}, might
benefit from the simplifications of a physics-oriented approach.

Engineering models of washboard formation range from coupled, damped
pendulum models~\cite{riley,riley_thesis} to full continuum
simulations of the deformable road surface~\cite{shoop}. Numerous
experimental studies have been undertaken, from laboratory scale
rigs~\cite{mather,riley,riley_thesis} to full scale road
tests~\cite{grau}.  In all cases, however, the engineering goal was to
understand all the complexities of the system and to mitigate or
eliminate the effect.  In contrast, we aim to understand the simplest
system that exhibits washboard road and to study it as a nonlinear,
pattern forming instability. Only a few theoretical studies of this
kind have appeared in the physics literature~\cite{both,mays}. In
addition to carrying out well instrumented laboratory-scale
experiments, we present here the first application of soft-particle
Discrete Element Method (DEM) simulations to this problem.  We also
present a simple theoretical treatment in order to gain some insight
into the fundamental mechanism of the instability and to understand
the scaling and important dimensionless groups.

\begin{figure}[b!]
  \centering
  \includegraphics*[width=65mm]{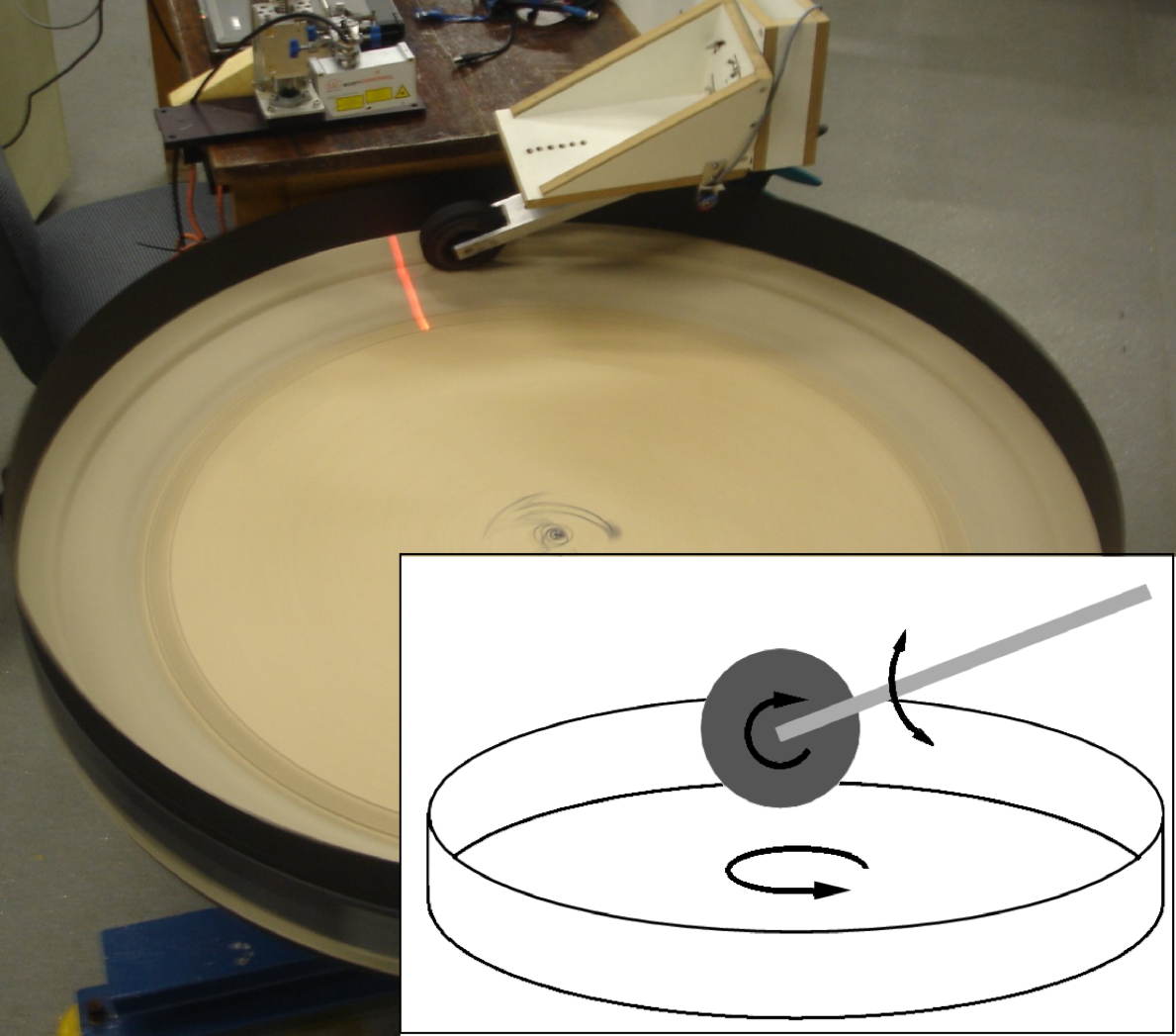}
  \caption{(Color online) Experimental setup.
    A bed of natural sand is laid on the circumference of a rotating
    table (1-m in diameter and rotation rate between 0.2Hz and 0.8Hz). A hard rubber wheel attached to an arm is free to bounce
    and roll on the granular bed.}
 \label{apparatus}
\end{figure}

The experimental setup is shown in Fig.~\ref{apparatus}. The road
consists of a deep layer of sand arranged on the circumference of a
1\,m diameter rotating table. We used natural, rough sand with a grain
diameter of 300 $\pm$ 100 $\mu$m and the bed was typically 50\,mm
deep. A 100\,mm diameter, 20\,mm thick hard rubber wheel was attached
to a 330\,mm long arm in the form of a lever.
The wheel rolled freely
on the sand bed as the table rotated at a constant  speed. No torque was applied to the wheel other than that produced by its contact with
the bed. The table was typically rotated at about 0.6\,Hz which
corresponds to a horizontal velocity of the wheel
$v\approx\mbox{2\,m/s}$.

A potentiometer was attached to the arm to record its angle, and its
output was digitized. Simultaneously, a commercial laser displacement
profiler\cite{profilometer} was used to record the bed shape.  Thus,
we could measure the wheel position and ripple shape with a vertical
accuracy of 1\,mm. Data acquisition was triggered by an optical sensor
fixed to the table, and several thousand vertical positions could be
measured in one table rotation.  During a typical run, the arm
inclination remained confined between $70^\circ$ and $90^\circ$, so
that the wheel motion was almost vertical.  The large angle implies
that the natural pendulum frequency of the arm plays no role, and the
wheel is merely restored by falling under its own weight.  The finite
circumference of the table effectively imposes periodic boundary
conditions on the ripple pattern, so that its wavelength is quantized
since a fully developed washboard pattern contains an integer number
of ripples around the table.

We also investigated the washboard formation in 2D DEM simulations.
The simulation considers individual deformable disks, rotating and
colliding with one another, subject to contact friction and gravity~\cite{Frenkel}.
We used the following physical parameters in the simulation: particle diameter 8\,mm,
mass 0.16\,g, spring constant $40~\mathrm{kNm^{-1}}$, coefficients of
restitution $0.5$ and friction $0.3$. In the simulation, the wheel
was treated like any other disk but its density was 1/5th that of the
other disks and its diameter was 12.5 times larger. A constant
horizontal velocity $v$ was imposed on the wheel, but it was free to
rotate and move vertically. The disks were made slightly poly-disperse
($\pm$20\% in diameter) in order to avoid crystallisation. In order to
mimic the experimental setup, the simulation was made periodic in the
horizontal direction.  25\,000 small disks were initialized at random
positions in the box, and then allowed to fall under gravity to settle
into the bed, resulting in a layer 20 diameters thick and 1500
diameters long.  The simulations were typically run for 500 passages
of the wheel.

Such 2D simulations cannot be expected to reproduce the experimental data quantitatively. The simulated grains are softer than natural sand and idealized as spherical, while the ratio of the diameter of the wheel to that of the grains is necessarily much larger in the experiment than in the simulation. In spite of these simplifications, the qualitative results of the DEM simulations offer unique new insights into the underlying mechanisms of the instability.

\begin{figure}[htbp]
 \centering
 \includegraphics*[width=\columnwidth]{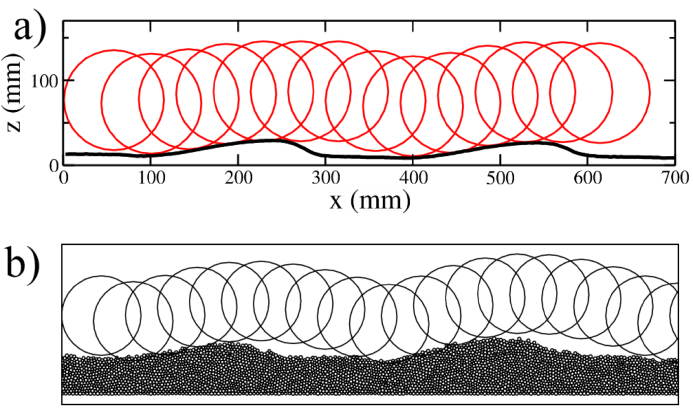}
 \caption{(Color online) Bed profile and wheel trajectory for typical
   washboard patterns obtained experimentally (a) and numerically (b).}
 \label{profiles}
\end{figure}

The wheel initially rolls smoothly on the flat granular surface. After
a few tens of passes, if the velocity is high enough, a small
localized ripple starts forming at some position. New ripples rapidly
grow from that location, downstream from the initial position.
Ripples then grow in height and eventually spread over the whole
circumference of the table (in the experiments) or the whole length of
the periodic box (in the simulation). A typical pattern is shown in
Fig.~\ref{profiles} which displays both experimental and simulation
data. There is an excellent qualitative agreement between the
experiment and simulation.

The ripples are strongly asymmetrical with the steeper face close to
the angle of repose. Individual ripples tend to be separated by flat
regions. The height of ripples ranges from a few mm up to 50\,mm as
the wavelength (or pitch) ranges from 50\,mm to 500\,mm.
Fig.~\ref{profiles} also shows the trajectory of the wheel. For
large $v$, the wheel becomes airborne near the crest of each ripple.
At lower $v$, however, a clear washboard pattern can form while the
wheel remains constantly in contact with the bed.

As is observed on actual roads~\cite{heath}, we found a critical value
of the driving velocity $v_c$ below which the bed remains flat. We
find experimentally that $v_c \simeq 1.5\,{\rm m s}^{-1}$, a
value quite similar to that found for real roads. The exact nature of
the bifurcation to the rippled state remains unclear, however. The bed
becomes extremely sensitive to small, but finite, perturbations for
$v$ near $v_c$, but it is difficult to establish whether $v_c$
represents the onset of a linear instability to infinitesimal
perturbations, or whether there is any velocity hysteresis near $v_c$.
Most previous
studies~\cite{heath,mather,riley,riley_thesis,grau,shoop} have only
examined the case of large, artificial initial perturbations and $ v
\gg v_c$.

Washboard patterns exist over a wide range of parameters. We varied
the bed thickness, the grain size and shape, the wheel size, shape and
mass. As long as the bed thickness was sufficient to supply enough
material for the ripples, in all cases the results were qualitatively
identical. All our observations are broadly consistent with previous
engineering studies~\cite{heath,mather,riley,riley_thesis,grau,shoop},
but have the advantage of a simpler suspension system and more modern
instrumentation for data acquisition.  Perhaps surprisingly, the size
and shape of the grains has no effect whatsoever on either the
wavelength or the amplitude of the ripples.  In experiments, we tried
two different natural sands with $d = 300\pm100\,\mathrm{\mu m}$ and
$3.0\pm0.8\,\mathrm{mm}$, and also replaced the sand with long grain
rice. In simulations, we halved the grain size and doubled the bed
thickness for the same size wheel. The ripple patterns were identical
up to small statistical fluctuations.  The mass of the wheel and its
suspension strongly affects the pattern.  Heavier wheels produce
larger amplitude ripples with shorter wavelengths. Wheel diameter,
however, seems to be unimportant, and this was tested by also using a non-rotating square ``wheel".  Varying the diameter of the wheel, keeping its mass constant in
the simulation,  leaves the pattern
unchanged.  The insensitivity of the pattern to the wheel diameter and grain size raises interesting questions about the scaling of the washboard
and are incorporated into the model discussed below.

\begin{figure}[htbp]
 \centering
 \includegraphics*[width=\columnwidth]{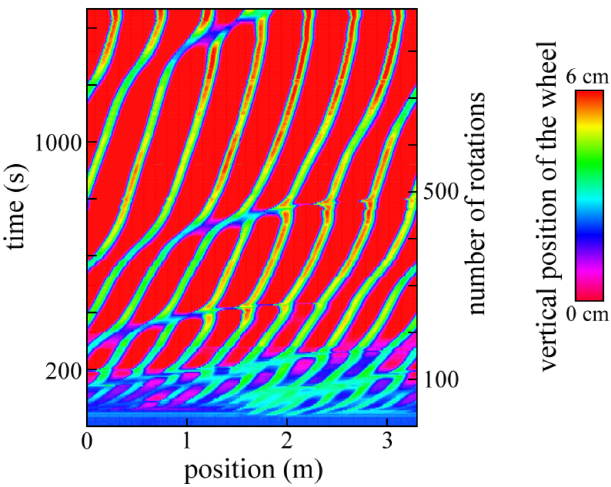}
 \caption{(Color online) An experimental space-time plot of the vertical position of
 the wheel for $v=\mbox{2 m s}^{-1}$.}
 \label{spacetime}
\end{figure}

\begin{figure}[htbp]
 \centering
 \includegraphics*[width=\columnwidth]{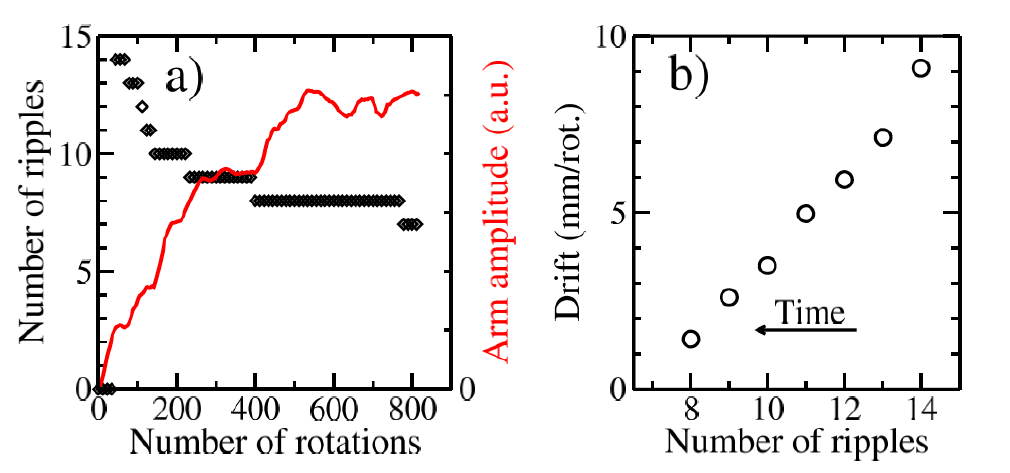}
 \caption{(Color online) Experimental ripple dynamics from
   Fig.~\ref{spacetime}. (a) Time evolution of the number of ripples
   (diamonds) and of the amplitude of the ripples (solid line). (b)
   Drift velocity in mm/rotation, as a function of the number of
   ripples.}
 \label{coarsen_drift}
\end{figure}

We find that both the amplitude and wavelength of the ripples grow
initially.  They remain roughly proportional to one another as they
evolve toward a saturated value which scales with the kinematic
length-scale $v^2/g$.  Thus, as they spread around the table, their
amplitude and wavelength increases and merging events take place.
Fig.~\ref{spacetime} is a typical space-time plot obtained
experimentally with $v = 2~{\rm m s}^{-1}$.  Our ripples always travel forwards,
whereas ripples on real roads, with driven wheels, have sometimes been
observed to travel in both directions~\cite{heath}.
Fig.~\ref{coarsen_drift}(a) shows the ripple evolution for the run shown in
Fig.~\ref{spacetime}. As the ripple amplitude saturates, the number of
ripples drops from 14 to 7. The amplitude increases abruptly each time
a ripple disappears. As the velocity is decreased, the ripples can
split and the amplitude and wavelength then decreases, but this is not
always observed.  Fig.~\ref{coarsen_drift}(b) shows that the ripple
drift velocity slows significantly as the wavelength increases, as one
would expect since the volume of individual ripples increases while the flux remains roughly constant. The DEM simulation qualitatively reproduces all of the behavior shown in Figs.~\ref{spacetime} and \ref{coarsen_drift}.

\begin{figure}[htbp]
  \centering
  \includegraphics*[width=\columnwidth]{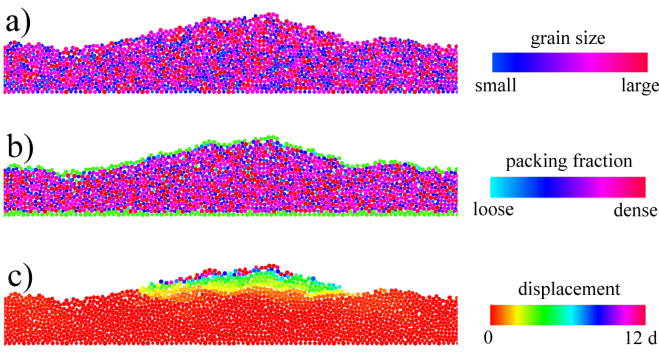}
  \caption{ (Color online) The internal structure of the ripples,
    using DEM simulation. (a) grain size, (b) local packing fraction,
    (c) displacement between two consecutive passes of the wheel.}
  \label{internal}
\end{figure}

Using DEM simulations, we can examine aspects of the internal
structure of the ripples which are difficult to access experimentally.
Fig.~\ref{internal}(a) shows that the slightly poly-disperse grains
remain well mixed inside the ripple. This demonstrates that size
segregation is not crucial to the formation of the ripples, although
it is almost certainly present on real roads~\cite{heath}.
Segregation nearly always occurs in granular systems with different particles and would probably occur in the DEM simulation if they
were run for much longer, but it is peripheral to the formation of ripples.

The local packing fraction of a pile can be computed from its
Vorono\"i tessellation. This algorithm partitions space into cells
corresponding to individual particles.  The area of each disk divided
by the area of its cell can be interpreted as a local packing
fraction. This is shown in Fig.~\ref{internal}(b).  The grains located
at the edge of the pile (in green) have an infinitely large, open
Vorono\"i cell for which no packing fraction can be defined. The
internal grains show no ripple-related structure in their packing,
although the overall packing is denser than the initial state of the
simulation. Thus, we conclude that varying compaction of the grains is
also not essential to the formation of ripples.
This contradicts a recent
model~\cite{both}, in which compaction played a central role.

Fig.~\ref{internal}(c) shows the displacement of the grains caused by
one pass of the wheel. The configuration is the state before the pass,
and the color shows the horizontal distance those grains are about to
travel. The displacement is localized to the crest of the ripples,
which move on a static bed. Although individual particle displacements
can be relatively large (up to 12 diameters), it typically takes 200
to 300 passes of the wheel for a ripple to travel a distance of one
wavelength.

While the dynamics of the wheel and its suspension are simple, the
behavior of granular materials is very poorly understood. There is no
continuum theory that can be reliably applied to this problem, though
phenomenological models exist which can be implemented with
large-scale finite-element codes~\cite{shoop}.  
On the other hand, existing simplified models are either based on compaction~\cite{both}, which we have shown to be unimportant, or are on rather {\it ad hoc} automata~\cite{mays}.

We now consider a model aimed at elucidating the physical scales of the ripple problem.  Stresses in a non-cohesive material are related to inertial forces or
gravitational forces, so the only dimensional parameters are the 2D
bed density $\rho$ (\textit{i.e.} the 3D density multiplied by the
width of the wheel) and the acceleration due to gravity $g$. The
experiments and simulations show that the particle size is not
important, if it is small enough compared to the wheel and ripples.
The primary dimensionless parameter needed to describe the material
is its approximate angle of repose $\theta_c$. If the wheel is
supported by static forces, the penetration of the wheel into the bed
is determined by its mass $m$, which gives rise to a penetration
length scale $L_1=\sqrt{{m}/{\rho}}$. There is a secondary dependence
on the radius of the wheel $R$ through the ratio $R/L_1$.  The
simulations show that this dependence is very weak.  The experiments
with the non-rotating square wheel, where no such length scale is
present, show that this is a secondary effect. The final significant
dimensional parameter is $v$, the horizontal speed of the wheel.
This gives rise to a second penetration length scale $L_2= mg/\rho
v^2$, if the wheel becomes supported by dynamic pressure. The ratio
$L_1/L_2$ behaves like a Froude number $\mbox{Fr} = (v^2/g)
\sqrt{\rho/ m}$. This is the only dimensionless group that can be
formed from $v$, $m$, $g$ and $\rho$, thus it should determine the
stability of the wheel/bed interaction.

We hypothesize that the instability results from a switch from
highly dissipative, plastic behavior at low speeds to weakly dissipative, dynamic
supporting forces at high speeds. In the absence of any special spring/dashpot
suspension, the saturated ripples involve ballistic trajectories, so
the natural length scale for the ripple wavelength $\lambda$ is the
length-scale $L=v^2/g$. Then $\lambda/L = f_\lambda(\mbox{Fr})$, where
$f_\lambda$ is a relatively weak function of $\mbox{Fr}$. We expect
the saturated amplitude of the ripples $A$ to be determined by
$\lambda$ and the angle of repose $\theta_c$, so that
$A/L=f_A(\mbox{Fr})$, with $f_A(\mbox{Fr}) \propto \tan \theta_c$.

The key element of a continuum theory to describe this system is the
function for the volume flux of sand $q$ which will be roughly
proportional to $zv$, where $z$ is depth the wheel penetrates into the
sand.  The flux will also depend on the slope angle and should diverge
as the surface angle approaches the angle of repose. In addition there
will be dependence on the local dimensionless groups $zg/v^2$ and
$Ng/\rho v^4$, where $N$ is the normal force. The penetration depth
$z$ will be determined by the forces acting on the wheel.  Thus, the key
to an accurate theory is to correctly model $N$. We propose
\begin{equation}
  N = \left\{
    \begin{array}{ll}
      0 & z<0 \\
      - z \delta \rho v^2 & z>0, \dot z <0\\
      - z \delta \rho v^2 + z^2 \mu \rho g & z>0, \dot z > 0\, ,
    \end{array}
  \right.
\end{equation}
where $\delta$ and $\mu$ are geometric factors that may depend
on $z/R$. This term is effective whether the wheel is moving up or
down and is conservative.  The second term is 
completely dissipative and represents the plastic response of the bed. Because of this strong piecewise nonlinearity, the system cannot be treated by linear
stability analysis.  The mean normal
force balances gravity and thus the displacement $z$ on a flat bed
will be of the order of $L_1/\mathrm{Fr}^2$. Thus, as $\mathrm{Fr}$
increases, the interaction with the bed becomes less dissipative and
the wheel can bounce upward
 from the bed. It is this bounce that drives the instability, because it allows a phase lag to develop between the volume flux $q$ and the driving force $N$, leading to a mutually reinforcing oscillation.

Washboard road will no doubt continue to annoy drivers for as long as
there are unpaved roads and wheels to roll over them.  We have re-examined this engineering problem from the perspective of basic nonlinear physics. We have argued that the appearance of the ripples at a critical speed should be regarded as an instability of the flat road.  Using simplified experiments and DEM simulations, we have shown that neither compaction nor segregation processes are responsible for the instability, contrary to some existing theories. Using very general dimensional analysis arguments, we have identified a candidate for the dimensionless parameter which controls the instability.  The piecewise form of the important nonlinearity in the normal force is such that the onset of the instability will not be amenable to  standard linear or weakly nonlinear stability analysis techniques.  Well above onset, we experimentally observed interesting ripple-merging and coarsening events, which are also reproduced by the DEM simulations.  We hope to use these insights to obtain a better understanding of this fascinating example of nonlinear pattern formation in granular media.

 NT was supported by the Newton Trust and the Swiss National Science
 Foundation. JNM was supported by the UK Engineering and Physical
 Sciences Research Council (GR/TO2416101). We are grateful to S.
 Dalziel, R. E. Ecke and T. B\"{o}rzs\"{o}nyi for experimental help,
 and to P. Richard for help with the Vorono\"i algorithm.


\begin{thebibliography}{99}
\bibitem{heath} For a review, see W. Heath and R. Robinson, {\it
 Transport and Road Research Laboratory}, Suppl. Report 610 (1980).

\bibitem{langlois} A. Valance and V. Langlois, {\it Eur. Phys. J. B}, {\bf 43}, 283 (2005).

\bibitem{herrmann} K. Kroy, G. Sauermann and Hans J. Herrmann, {\it Phys. Rev. Lett.}, {\bf 88}, 054301 (2002).

\bibitem{mather} K. B. Mather, {\it Civ. Eng. Pub. Works Rev.}, {\bf
 57}, 617 (1963), and {\it Civ. Eng. Pub. Works Rev.}, {\bf 57},
 781 (1963).

\bibitem{riley} J. G. Riley and R. B. Furry, {\it Highway Research
 Record}, {\bf 438}, 54 (1973).

\bibitem{riley_thesis} J. G. Riley, Ph.D thesis, Cornell University
 (1971).

\bibitem{grau} R. W. Grau and L. B. Della-Moretta,
\emph{Trans. Res. Rec.} \textbf{1291}, Vol. 2, 313 (1991).

\bibitem{shoop} S. Shoop, R. Haehnel, V. Janoo, D. Harjes and R.
  Liston, {\it J. Geotech. Geoenviron. Eng.},
  \textbf{132},  852 (2006).
  

\bibitem{mays} D. C. Mays and B. A. Faybishenko, {\it Complexity},
 {\bf 5}, 51 (2000). 

\bibitem{both} J. A. Both, D. C. Hong and D. A. Kurtze, {\it Physica
 A}, {\bf 301}, 545 (2001).

\bibitem{profilometer} Micro-Epsilon LLT2800--100 2D Laser displacement
  measuring system.

\bibitem{Frenkel} D. Frenkel and B. Smit, Understanding Molecular Simulation, {\it Academic Press, San Diego},
(1996).

\end{thebibliography}
\end{document}